\documentclass[twocolumn,showpacs,preprintnumbers,amsmath,amssymb]{revtex4}
\usepackage{dcolumn}
\usepackage{bm}
\usepackage[dvipdf]{graphicx}
\DeclareGraphicsExtensions{.jpg,.pdf,.mps,.png,.pdf,.ps,.pdf}

\newcommand{\be}{\begin{equation}}
\newcommand{\ee}{\end{equation}}
\newcommand{\bea}{\begin{eqnarray}}
\newcommand{\eea}{\end{eqnarray}}
\newcommand{\ii}{\text{i}}

\newcommand{\avg}[1]{\left\langle{#1}\right\rangle}
\newcommand{\bsy}[1]{\boldsymbol{#1}}
 
\newcommand{\sign}{\textrm{sign}\ }
\newcommand{\beas}{\begin{eqnarray*}}
\newcommand{\eeas}{\end{eqnarray*}}
\def\sign{\hbox{sign}\,}

\newcommand{\BE}{\begin{eqnarray}}
\newcommand{\EE}{\end{eqnarray}}
\newcommand{\BEn}{\begin{eqnarray*}}
\newcommand{\EEn}{\end{eqnarray*}}
\newcommand{\barr}{\begin{array}}
\newcommand{\earr}{\end{array}}

\newcommand{\bit}{\begin{itemize}}
\newcommand{\eit}{\end{itemize}}
\newcommand{\bc}{\begin{center}}
\newcommand{\ec}{\end{center}}
\newcommand{\ben}{\begin{enumerate}}
\newcommand{\een}{\end{enumerate}}

\newcommand{\ovl}{\overline}

\newcommand{\Tr}{\text{Tr}}
\newcommand{\Avg}{\avg}

\newcommand{\om}{\omega}

\renewcommand{\l}{\left}
\renewcommand{\r}{\right}
\begin{document}
\newcommand{\hT}{ {\cal T} }
\newcommand{\hB}[2]{ {\cal B} \left[ {#1} \cdot {#2} \right]}
\title{Multi-asset minority games} \author{G. Bianconi$^{1}$, A. De
  Martino$^2$, F. F. Ferreira$^3$ and M. Marsili$^1$}
   \affiliation{$^1$The Abdus Salam ICTP, Strada
  Costiera 14, 34014 Trieste, Italy} \affiliation{$^2$CNR-ISC,
  INFM-SMC and Dipartimento di Fisica, Universit\`a di Roma ``La
  Sapienza'', p.le A. Moro 2, 00185 Roma, Italy} \affiliation{$^3$ EACH, Universidade De S\~ao Paulo, Av. Arlindo Bétio 1000,03828-080 S\~ao Paulo,Brazil}
\begin{abstract}
  We study analytically and numerically Minority Games in which agents
  may invest in different assets (or markets), considering both the
  canonical and the grand-canonical versions. We find that the
  likelihood of agents trading in a given asset depends on the
  relative amount of information available in that market. More
  specifically, in the canonical game players play preferentially in the stock
  with less information. The same holds in the grand canonical game when
  agents have positive incentives to trade, whereas when agents payoff are solely
  related to their speculative ability they
  display a larger propensity to invest in the information-rich
  asset. Furthermore, in this model one finds a globally predictable
  phase with broken ergodicity.
\end{abstract}
\pacs{: } \maketitle
\section{Introduction}
The study of systems of heterogeneous adaptive agents through
Minority Games (MGs) \cite{ElFarol,CZ} has attracted much interest
from statistical physicists. Despite the simplicity of the
interactions between agents, these models generate rich static and
dynamical structures which can often be well understood at the
mathematical level through the use of spin-glass techniques
\cite{book,cool}.  While the MG has found applications in
different types of problems (see for example \cite{traffic}), it
was originally designed to address the issue of how the
microscopic behavior of traders -- speculators in particular  --
may give rise to the anomalous global fluctuation phenomena
observed empirically in financial markets. In this respect the
most successful version of the MG has perhaps been the
grand-canonical MG \cite{Johnson,GCMG}, in which traders may
abstain from investing, so that the traded volume fluctuates in
time.

The core of MGs is the assumption that traders react to the
receipt of an information pattern (be it exogenous or endogenous)
by taking a simple trading decision such as buying or selling. The
key control parameter is the ratio between the number of traders
and the `complexity' of the information space, measured by the
number of possible patterns. In general, when this ratio exceeds a
certain threshold MGs undergo a phase transition to a
macroscopically efficient state where it is not possible to
predict statistically whether a certain decision will be fruitful
or not based on the received information alone.

Real markets are typically formed by different assets and are
characterized by non trivial correlations \cite{Mantegna,Potters,Kertesz}.
These correlations arise from the underlying behavior of the
economics (the fundamentals) but they are also ``dressed'' by the
effect of financial trading. In this paper, we will use the
Minority Game in order to investigate how speculative trading
affects the different assets in a market. Versions of MG where
agents are engaged in different contexts have already been
introduced and studied \cite{Rodgers,Chau}. More precisely, we shall
investigate how speculative trading contributes to financial
correlations, and how speculators distribute their trading volume
depending on the information content of the different asset
markets. 

Our first result is that speculative trading does not contribute
in a sensible manner to financial correlations, and if it does, it
likely contributes a negative correlation. The reason is that,
within the schematic picture of the MG, speculators are uniquely
driven by profit considerations and totally disregard risk. The
same cannot be said for strategies on lower frequencies (buy and
hold) where risk minimization of the portfolio becomes important.

Our second main conclusion is that, when there are positive
incentives to trade, speculators invest preferentially on the
asset with the smallest information content. This apparently
paradoxical conclusion is reverted when speculators have no
incentive to trade, other than making a profit. This is due to the
fact that speculators, when they are forced to trade also
contribute to information asymmetries. 

Finally, with respect to the usual classification in phases of the MG,
we find a considerably richer phase diagram where different components of the
market may be in different phases. These conclusions are derived for the case of a market
composed of two assets, which allows for a simpler treatment and
provides a more transparent picture. Their validity can be
extended in straightforward ways to the case of markets with a
generic number of assets.

The paper is articulated in three parts. Section \ref{sect:MG} is
dedicated to the study of a canonical MG where agents can choose
on which of two assets to invest. In Section \ref{sect:GCMG} we
discuss the grand-canonical version of this model, where agents
are also allowed to refrain from investing. Finally, we formulate
our conclusions in Sec. \ref{sect:concl}. The mathematical
analysis of the models we consider is a generalization of
calculations abundantly discussed in the literature (see
\cite{book,cool} for recent reviews). We therefore won't go into
the details, limiting ourselves to stressing the main differences
with the standard cases.

\section{Canonical Minority Game with two assets}
\label{sect:MG}

We consider the case of a market with two assets
$\gamma\in\{-1,1\}$ and $N$ agents. At each time step $t$,
agents receive two information patterns
$\mu_\gamma\in\{1,\ldots,P_\gamma\}$, chosen at random and
independently with uniform probability. It is assumed that
$P_\gamma$ scales linearly with $N$, and their ratio is denoted by
$\alpha_\gamma=P_\gamma/N$. Every agent $i$ disposes of two
``strategies'' $\bsy{a}_{i\gamma}=\{a_{i\gamma}^{\mu_\gamma}\}$
(one for each asset), that prescribe a binary action
$a_{i\gamma}^{\mu_\gamma}\in\{-1,1\}$ (buy/sell) for each possible
information pattern. Each component $a_{i\gamma}^{\mu_\gamma}$ is
assumed to be selected randomly and independently with uniform
probability and is kept fixed throughout the game. Traders keep
tracks of their performance in the different markets through a
score function $U_{i\gamma}(t)$ which is updated with the
following rule:

\be\label{learn1}
U_{i\gamma}(t+1)=U_{i\gamma}(t)-a^{\mu_\gamma(t)}_{i\gamma}
A_\gamma(t)
\ee

\noindent
where

\be
A_\gamma(t)=\sum_{j=1}^N
a^{\mu_\gamma(t)}_{j\gamma}\delta_{s_j(t),\gamma}
\label{Agamma}
\ee

\noindent represents the `excess demand' or the total bid on
market $\gamma$ (the factor $1/\sqrt{N}$ appears here for
mathematical convenience) and is usually taken as a proxy of (log)
returns, i.e.
$\log p_\gamma(t+1)=\log p_\gamma(t)+\lambda A_\gamma(t)$.
The Ising variable

\be
s_i(t)=\sign\left[U_{i,+1}(t)-U_{i,-1}(t)\right] \label{si}
\ee

\noindent
indicates the asset in which player $i$ invests at time $t$,
which is simply the one with the largest cumulated score.
It is the minus sign on the right-hand side of (\ref{learn1})
that enforces the minority-wins rule in both markets:
Agents will invest in that market where their strategy provides a
larger payoff $-a_{i\gamma}^{\mu_\gamma(t)}A_\gamma(t)$
(or a smaller loss).

It is possible to characterize the asymptotic behaviour of the
multi-agent system (\ref{learn1}) with a few macroscopic
observables, such as the predictability $H$ and the volatility $\sigma^2$,
defined respectively as \cite{H}
\begin{gather}
  H=\sum_{\gamma\in\{-1,1\}}\frac{1}{NP_\gamma}\sum_{\mu_\gamma=1}^{P_\gamma}
  \avg{A_\gamma|\mu_\gamma}^2=H_++H_-\label{acca}\\
  \sigma^2=\frac{1}{N}\sum_\gamma\avg{A_\gamma^2}=
  \sigma^2_++\sigma^2_-\label{s2}
\end{gather}
with $\avg{\cdot}$ and $\avg{\cdot|\mu_\gamma}$ denoting time
averages in the steady state, the latter conditioned on the
occurrence of the information pattern $\mu_\gamma$. Besides these, in the
present case, it is also important
to study the relative propensity of traders to invest in a given
market, namely

\be
m=\frac{1}{N}\sum_{i=1}^N\avg{s_i}
\label{parapa}
\ee

\noindent
A positive (resp. negative) $m$ indicates that
agents invest preferentially in asset $+1$ (resp. $-1$).

\begin{figure}
\includegraphics*[width = 8cm]{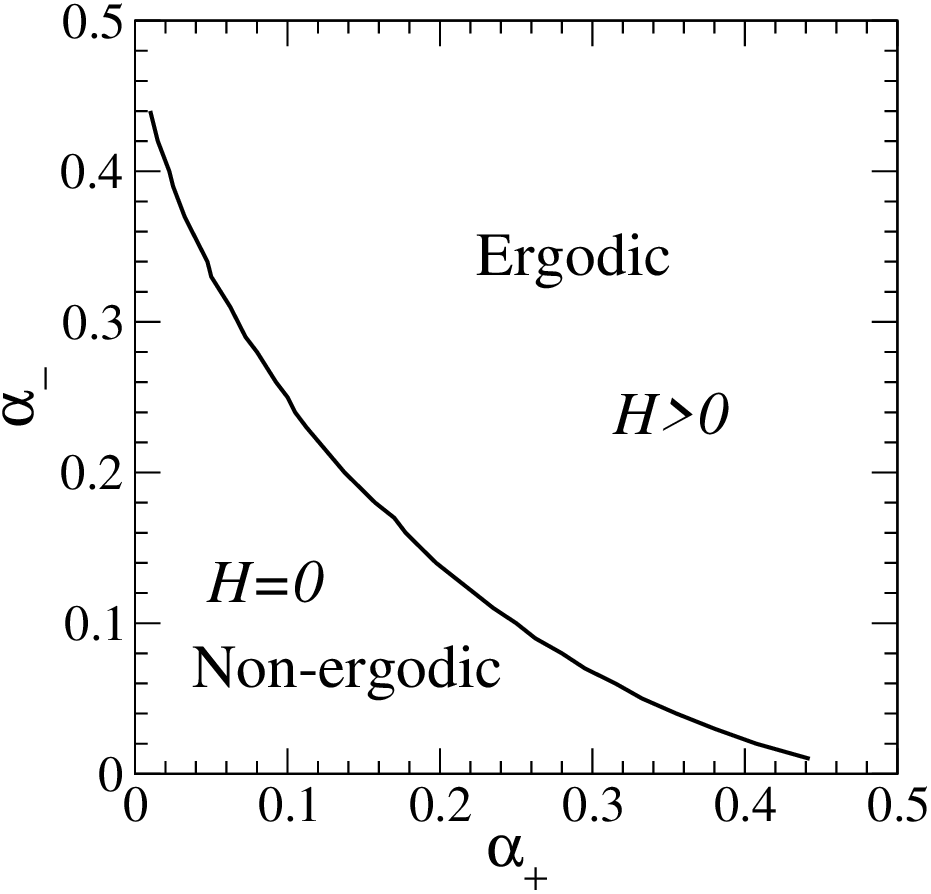}
\caption{Analytical phase diagram of the canonical two-asset Minority
Game in the $(\alpha_+,\alpha_-)$ plane.}
\label{Phase_diagram.fig}
\end{figure}

It is clear, already at this stage, that if no {\em a priori}
correlation is postulated between the news arrival processes on the
two assets $\mu_{\pm}(t)$ or between the strategies adopted by agents
in the two markets, no correlation is created by agents.
Indeed
\begin{equation}\label{Acor}
  \avg{A_+ A_-}=\frac{1}{N}\sum_{i,j}\left\langle
  a_{i,+}^{\mu_+}a_{i,-}^{\mu_-}\frac{1+s_i}{2}\frac{1-s_j}{2}
  \right\rangle
\end{equation}
and we know \cite{MC01} that dynamical variables $U_{i\gamma}(t)$ evolve
on timescales much longer (of order $P_\gamma$) than that over which
$\mu_\gamma$ changes. Hence we can safely assume that the distribution of
$s_i$ in Eq. (\ref{Acor}) is independent of $\mu_{\pm}$, which allows to
factor the average $\langle a_{i,+}^{\mu_+}a_{i,-}^{\mu_-}\rangle
=\langle a_{i,+}^{\mu_+}\rangle\langle a_{i,-}^{\mu_-}\rangle$ over
the independent information arrival processes $\mu_{\pm}(t)$. Given that
$\langle a_{i,\pm}^{\mu_{\pm}}\approx 0$ we conclude that
$\avg{A_+ A_-}\approx 0$ also. The reason for this is that traders behavior is
aimed at detecting excess returns in the market with no consideration about
the correlation among assets.
The quantities defined above can be obtained both numerically and
analytically (in the limit $N\to\infty$) as functions of $\alpha_+$ and
$\alpha_-$.  The phase structure of
the model is displayed in Fig. \ref{Phase_diagram.fig}.
The $(\alpha_+,\alpha_-)$ plane is divided in two regions separated by
a critical line. In the ergodic regime, the system produces
exploitable information, i.e. $H>0$, and the dynamics is ergodic, that
is the steady state turns out to be independent of the initialization
$U_{i\gamma}(0)$ of (\ref{learn1}). Below the critical line, instead,
different initial conditions lead to steady states with different
macroscopic properties (e.g. different volatility). In this region traders manage
to wash out the information and the system is unpredictable
($H=0$). This scenario essentially reproduces the standard MG phase
transition picture.  The model can be solved analytically in two
complementary ways and in both cases calculations are a
straightforward generalization of those valid for the single-asset
case. The static approach relies on the fact that the stationary
state is described by the minima of the random function

\be
H=\sum_{\gamma\in\{-1,1\}}\frac{1}{NP_\gamma}\sum_{\mu_\gamma=1}^{P_\gamma}\l[
\sum_{j=1}^N a_{j\gamma}^{\mu_\gamma}\frac{1+\gamma m_j}{2}\r]^2
\ee
over the variables $-1\leq m_i=\avg{s_i}\leq 1$.
$H$ coincides with the
predictability in the steady state, which implies
that speculators make the market as unpredictable as possible.
The statistical mechanics approach proceeds by studying the properties of a
system of soft spins $m_i$ with Hamiltonian $H$ at a fictitious inverse
temperature $\beta$ in the limit
$N\to\infty$.
The relevant order parameter
is the overlap $Q_{ab}=(1/N)\sum_i m_{ia}m_{ib}$
between different minima $a$ and $b$, which takes
the replica-symmetric form $Q_{ab}=q+(Q-q)\delta_{ab}$. Phases where
the minimum is unique, corresponding to $H>0$, are described by taking
$Q\to q$ (evidently) and $\chi=\beta(Q-q)$ finite in the limit $\beta\to\infty$.
The condition $\chi\to\infty$ signals the phase transition to the unpredictable
phase with $H=0$.

The dynamical approach employs path-integrals to transform the $N$
coupled single-agent processes for the variable
$y_i(t)=U_{i,+1}(t)-U_{i,-1}(t)$ into a single
stochastic process equivalent to the original $N$-agent system in
the limit $N\to\infty$ \cite{cool}. The calculation is greatly
simplified if one studies the `batch' version
\cite{CoolHeim}, which roughly corresponds to a time re-scaling
$t\to \tau=t/N$ and, apart from the value of $\sigma^2$, has the same
collective behavior.
In this case, the effective process has the form
\begin{equation}
y(\tau+1)=y(\tau)-\sum_{\gamma,\tau'}\l[\boldsymbol{1}+
\frac{\boldsymbol{G}}{2\alpha_\gamma}\r]^{-1}(\tau,\tau')
\frac{\gamma+s(\tau')}{2}+z(\tau)
\end{equation}
where $z(\tau)$ is a zero-average Gaussian
noise $z(\tau)$ with correlation matrix $\avg{z(\tau)z(\tau')}=\Lambda(\tau,
\tau')$
with
\be
\bsy{\Lambda}=\sum_\gamma\l[\l(\boldsymbol{1}+
\frac{\boldsymbol{G}}{2\alpha_\gamma}\r)^{-1}\l(n_\gamma\boldsymbol{D}_\gamma\r)
\l(\boldsymbol{1}+ \frac{\boldsymbol{G}^T}{2\alpha_\gamma}\r)^{-1}\r]
\label{nv}
\ee
where
\begin{gather}
D_\gamma(\tau,\tau')=\frac{1}{4}\l[1+\gamma m(\tau)+\gamma m(\tau')+
C(\tau,\tau')\r]\\
m(\tau)=\avg{s(\tau)}~~~~~~~C(\tau,\tau')=\avg{s(\tau)s(\tau')}
\end{gather}
while $G(\tau,\tau')=\avg{\frac{\partial s(\tau)}{\partial h(\tau')}}$ denotes the
response to an infinitesimal probing field $h(\tau)$.  Both $H$ and
$\sigma^2$ can be obtained from the asymptotic study of
$\Lambda(\tau,\tau)$. Ergodic steady states, where $C(\tau,\tau')=c(\tau-\tau')$
and $G(\tau,\tau')=g(\tau-\tau')$, can be described in terms of three variables
only, namely the ``magnetization''
$m=\lim_{\tau\to\infty}\frac{1}{\tau}\sum_{\tau'}m(\tau')$, the persistent
autocorrelation $q=\lim_{\tau\to\infty}\frac{1}{\tau}\sum_{\tau'}c(\tau')$
and the susceptibility $\chi=\lim_{\tau\to\infty}\sum_{\tau'}G(\tau')$,
for which one derives closed equations that can be solved numerically.
The results for $m,~q$ and $\chi$ coincide with those obtained in the
static approach, thus providing a dynamic interpretation for these quantities.
It turns out that $\chi$ diverges as the line in Fig. \ref{Phase_diagram.fig} is
approached from above, signalling ergodicity breaking and the onset of a phase
in which the steady state depends on the initial conditions of
the dynamics. We find
\be
H=\sum_{\gamma} \frac{\alpha_{\gamma}^2 (1+2\gamma
m+q)^2}{[2\alpha_{\gamma}+\chi]^2}
\ee
which implies $H=0$ in the non-ergodic phase.

\begin{figure}
\includegraphics*[width = 8cm]{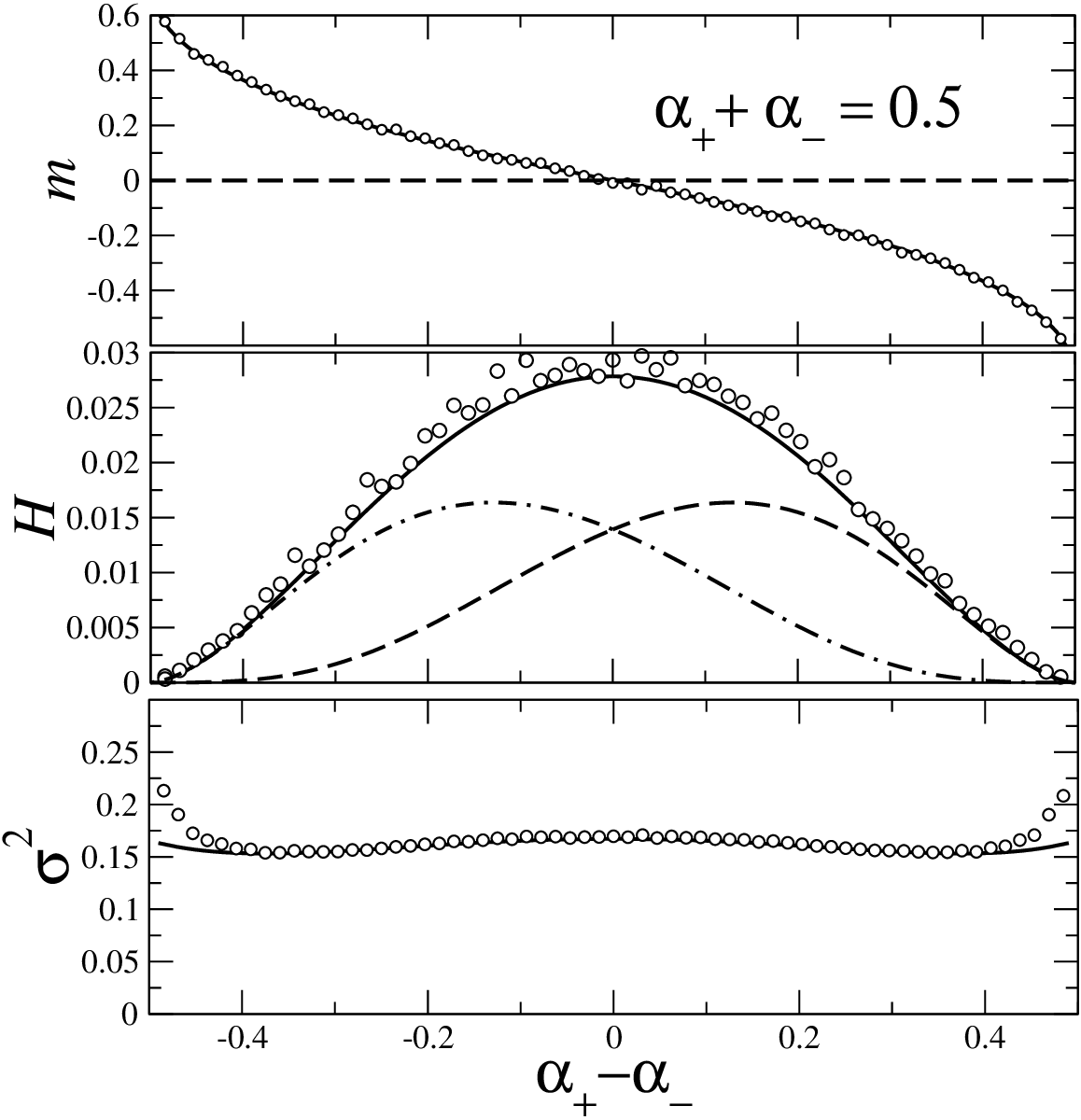}
\caption{Behavior of $m$ (top), $H$ (middle) and $\sigma^2$ (bottom)
versus $\alpha_+-\alpha_-$ for $\alpha_++\alpha_-=0.5$. Markers
correspond to simulations with $N=256$ agents, averaged over 200
disorder samples per point. Lines are analytical results (see
Appendices A and B for details). In the middle panel, the dashed line
corresponds to $H_+$ and the dot-dashed line corresponds to $H_-$.}
\label{P_tot0.5}
\end{figure}

For the volatility (of the original on-line case), one
obtains instead the approximate expression
\be
\sigma^2=H+\frac{1-q}{2}
\ee
which is very accurate in the ergodic phase.
The behaviour of these quantities
along a cut $\alpha_++\alpha_-=$ constant in the ergodic phase is
reported in Fig. \ref{P_tot0.5} together with that of the order
parameter $m$.

A na\"\i ve argument would suggest that agents are attracted by information
rich markets. Instead one sees that, in a range of parameters, agents play
preferentially in the market with smaller information complexity
$\alpha_\gamma$ and with the smallest information
content $H_\gamma$.
For all those traders with $|m_i|<1$, the conditions for the minimum of $H$
give
\begin{equation}\label{eqmi}
  m_i=-\avg{a_{i+}^{\mu_+}A_+^{(-i)}}+\avg{a_{i-}^{\mu_-}A_-^{(-i)}}
\end{equation}
where $A_\gamma^{(-i)}$ stands for the contribution to $A_\gamma$ of all traders
except $i$. Hence $m_i$ equals the difference in the payoffs of agent $i$ against
all other traders and this relation means that if $m_i>0$ then agent $i$
invests preferentially in asset $+$ because that is more
convenient. Therefore, Fig. \ref{P_tot0.5} implies that the
relation between payoffs and information is less obvious than the na\"\i ve
argument above suggests.

This somewhat paradoxical result is due to the fact that
agents are constrained to trade in one of the two markets.
Rather than
seeking the most profitable asset, agents escape the asset where
their loss is largest.

\section{Grand Canonical Minority Game with two assets}
\label{sect:GCMG}

In the grand-canonical framework players have the option not to
play if their expected payoff doesn't beat a pre-determined
benchmark (which represents for instance a fixed interest rate or an
incentive to enter the market) \cite{GCMG}. As in the
previous case, we consider two assets or markets, tagged by
$\gamma\in\{-1,1\}$ as before. Each trader disposes of one
quenched random strategy
$\bsy{a}_{i\gamma}=\{a_{i\gamma}^{\mu_\gamma}\}$ per asset, which
prescribes an action $a_{i\gamma}^{\mu_\gamma}\in\{-1,1\}$ for
each possible information pattern $\mu_\gamma\in\{1,\ldots,P_\gamma\}$.
Again $\mu_\gamma(t)$ are
chosen at random independently for all $t$ and $\gamma=\pm 1$.
As in the one-asset grand-canonical MG,
it is necessary to introduce a certain number
of traders -- so-called producers -- who invest at every
time step no matter what. These can be regarded as traders with
a fixed strategy $b_{i\gamma}^{\mu_\gamma}=\pm 1$.
The number of producers in market
$\gamma$ shall be denoted by $N_{p,\gamma}$ and their
aggregate contribution to $A_\gamma(t)$ by
$B_\gamma^{\mu_\gamma(t)}=\sum_{i=1}^{N_p} b_{i,\gamma}^{\mu_\gamma(t)}$.
Therefore Eq. (\ref{Agamma}) becomes

\be
A_\gamma(t)=\sum_{j=1}^N
a^{\mu_\gamma(t)}_{j\gamma}\delta_{s_j(t),\gamma}+B_\gamma^{\mu_\gamma(t)}
\label{ABgamma}
\ee
The rest of the traders, the speculators, have an adaptive behavior which is
again governed by the dynamics (\ref{learn1}) but now agents can also decide
not to trade. This choice is equivalent to trading
in a fictitious $\gamma=0$ ``asset'' whose cumulated score is
$U_{i,0}(t)=\epsilon t$. More precisely
\begin{equation}\label{sigc}
  s_i(t)={\rm arg}\max_{\gamma\in \{0,\pm 1\}} U_{i,\gamma}(t)
\end{equation}
The $\gamma=0$ choice represents a fixed benchmark with
a constant payoff.
By Eq. (\ref{sigc}) traders invest in $\gamma\not =0$ assets only if their
score exceeds that of the benchmark, i.e. if the corresponding score
$U_{i\gamma}$ grows at least as $\epsilon t$.
Notice that agents are allowed to invest
in at most one asset. If agents were allowed to invest in both $\gamma=\pm 1$
assets if $U_{i,\pm}(t)>U_{i,0}(t)$ then it is easy to see that the model
becomes equivalent to two un-coupled GCMGs.

The arguments of the previous section show that also in this case
no significant correlation between assets is introduced by the behavior of
speculators.
Again the collective properties of the stationary state can be characterized
by the predictability $H$, Eq. (\ref{acca}), the
volatility $\sigma^2$, Eq. (\ref{s2}) and the ``magnetization" $m$ of
Eq. (\ref{parapa}). These
parameters can be studied as before upon varying the parameters
$\alpha^\gamma=P_\gamma/N$ and $\epsilon$. We also introduce the relative number
of producers $n_p=N_{p,\gamma}/P_\gamma$, which for simplicity is
assumed to be the same for both assets. Notice that for $n_p=0$ and
$\epsilon\to -\infty$ we recover the model of the previous section where
there are no producers and speculators are forced to trade.

\begin{figure}
  \includegraphics*[width = 8cm]{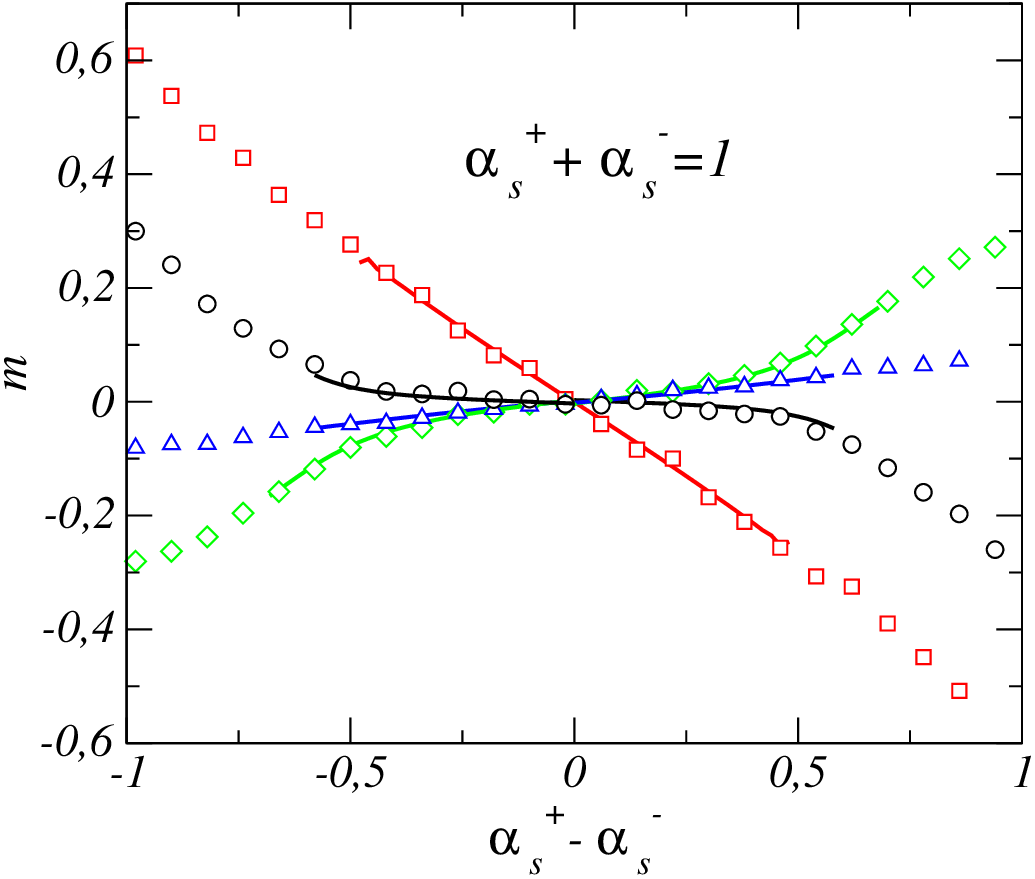}
\caption{Behavior of $m$ versus $\alpha_s^+-\alpha_s^-$ for
  $\alpha_s^++\alpha_s^-=1$ and flat initial conditions. Markers
  correspond to simulations with $N_s=200$ speculators, averaged over
  200 disorder samples per point.  Lines are analytical results
  (interrupted when the non-ergodic phase is met, see phase
  diagram). Other parameters are as follows: $\epsilon=0.1$, $n_p=1$
  ($\diamond$); $\epsilon=0.1$, $n_p=0.1$ ($\bigtriangleup$);
  $\epsilon=-0.1$, $n_p=1$ ($\bigcirc$); $\epsilon=-0.1$, $n_p=0.1$
  ($\square$).}
\label{mgrcn}
\end{figure}

We focus first on $m$ (see Fig. \ref{mgrcn}).
One sees that when traders have positive incentives to trade ($\epsilon<0$)
the market behaves as in the previous section, with speculators investing
preferentially in the asset with less
information. This tendency becomes less pronounced the larger is
$n_p$, which is reasonable in view of our discussion above,
because then the game becomes more and more
profitable for speculators.

This scenario is qualitatively reproduced at all
$\epsilon<0$ and it changes drastically as soon as $\epsilon>0$.
In this case, traders concentrate most of their investments into
the information-rich asset even if $n_p$ is very low. The
fact that traders can refrain from investing implies that trading
is dominated by gain seeking rather than escaping losses.

The theory for this case is slightly more involved than for the
canonical model.  On the static side, the Hamiltonian is now
\be
H_\epsilon=\sum_{\gamma =\pm 1}\frac{1}{N P_\gamma}
\sum_{\mu_{\gamma}=1}^{P_{\gamma}} \l[ \sum_{j=1}^{N}
a_{j\gamma}^{\mu_\gamma}
\pi_j^\gamma+B_\gamma^{\mu_\gamma}\r]^2+\frac{2\epsilon}{N}
\sum_{j=1}^{N}\pi_j^0\label{H.eq}
\ee
where $\pi_i^\gamma=\avg{\delta_{s_i,\gamma}}$ denotes the frequency with which
agent $i$ invests in asset $\gamma=0,\pm 1$ in the steady state. Notice that
$H_0=H$ is the predictability. As before, it is necessary to introduce a
fictitious temperature $\beta>0$ and turn to the replica trick to
analyze the minima of $H_\epsilon$ over $\{\pi_i^\gamma\}$ considering the
limit $\beta\to\infty$. The main difference with the canonical model
lies in the fact that one must now consider an overlap order parameter
per asset, namely
$Q_{ab}^\gamma=(1/N)\sum_{i=1}^{N}\pi_{ia}^\gamma \pi_{ib}^\gamma$
($a,b=1,\ldots,r$, $\gamma=\pm 1$) and, in the replica-symmetric Ansatz, one
`susceptibility' per asset, that is
$\chi^\gamma=\beta(Q^\gamma-q^\gamma)$.

Again these quantities can be given a dynamic interpretation with the
generating function approach \cite{cool}. This approach leads, in the
batch approximation, to two effective processes (one per asset),
namely
\begin{multline}
U_\gamma(\tau+1)=U_\gamma(\tau)-\epsilon +
z_\gamma(\tau)\sqrt{1+\alpha_\gamma n_p}\\-(1+\alpha_\gamma n_p)
\sum_{\tau'}\l[\bsy{1}+\lambda_\gamma\bsy{G}_\gamma\r]^{-1}(\tau,\tau')
\phi_\gamma(\tau')
\end{multline}
where $\lambda_\gamma=\frac{1+\alpha_\gamma n_p}{\alpha_\gamma}$
and the noise correlations are described by the matrices
$\Lambda_\gamma(\tau,\tau')=\avg{z_\gamma(\tau)z_\gamma(\tau')}$ with
\begin{gather}
\bsy{\Lambda}_\gamma=\l[\l(\bsy{1}+\lambda_\gamma\bsy{G}_\gamma\r)^{-1}
\l(\lambda_\gamma\bsy{C}_\gamma\r)\l(\bsy{1}+\lambda_\gamma
\bsy{G}_\gamma^T\r)^{-1}\r]\\
C_\gamma(\tau,\tau')=\avg{\phi_\gamma(\tau)\phi_\gamma(\tau')}~~~~~
G_\gamma(\tau,\tau')=\avg{\frac{\partial\phi_\gamma(\tau)}{\partial
h_\gamma(\tau')}}
\end{gather}
In order to characterize time-translation invariant and ergodic steady
states four quantities are now required, namely two persistent
autocorrelations
$q_\gamma=\lim_{\tau,\tau_0\to\infty}\frac{1}{\tau}\sum_{\tau'=1}^\tau
C_\gamma(\tau_0,\tau_0+\tau')$ and two
susceptibilities $\chi_\gamma=\lim_{\tau,\tau_0\to\infty}
\sum_{\tau'=1}^\tau G_\gamma(\tau_0,\tau_0+\tau')$,
$\gamma\in\{-1,1\}$. For these, one obtains equations that can be
solved numerically and the quantity $m$ can be written in terms of the
$q_\gamma$'s and the $\chi_\gamma$'s. Now ergodicity breaking is
connected to the divergence of at least one of the susceptibilities.

\begin{figure}
  \includegraphics*[width = 8cm]{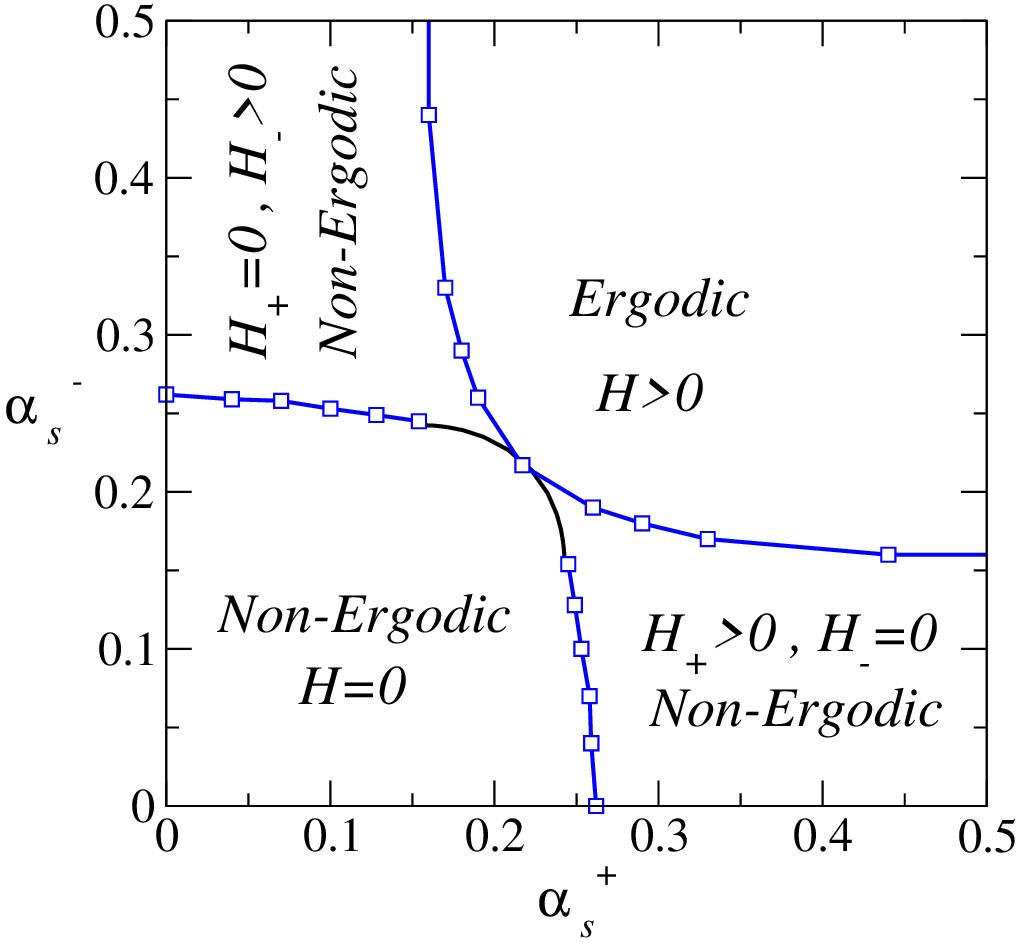}
\caption{Phase diagram of the $\epsilon=0$, $n_p=1$ grand-canonical
two-asset Minority Game in the $(\alpha_s^+,\alpha_s^-)$ plane. The
continuous line is analytical, the other phase boundaries are obtained
from numerical simulations (averages over 100 disorder samples per
point).}
\label{pdgc}
\end{figure}

The behavior of the model is considerably richer than in the previous case:
For $\epsilon\not =0$ we find that $H_\epsilon$ has a unique non-degenerate
minimum and both $\chi^\gamma$'s are finite. The case $\epsilon=0$ is peculiar
as it marks the boundary between two different behaviors $\epsilon<0$
and $\epsilon>0$. For $\epsilon=0$ and $\alpha_{\pm}$ large enough,
both markets are predictable ($H_0>0$) and the susceptibility is finite.
However, one of the susceptibilities diverges while
the other stays finite for lower values of $\alpha_\pm$.
This signals the onset of a phase where
one of the markets is unpredictable while still $H_0>0$, a situation
with particularly striking dynamical consequences.
As a result, the phase structure of this model is rather complex (see
Fig. \ref{pdgc}).
We have been unable to obtain analytical lines for the complete phase
structure at $\epsilon=0$. The phase boundary separating the region
with $H=0$ from that with $H>0$ has been calculated assuming that both
susceptibilities diverge keeping a finite ratio $\chi_+/\chi_-$. The
phase boundary of the non-ergodic region (which would correspond to
the divergence of just one of the susceptibilities) has been instead
estimated from numerical simulations and the corresponding lines must
be considered a crude approximation.

\begin{figure}
  \includegraphics*[width = 8cm]{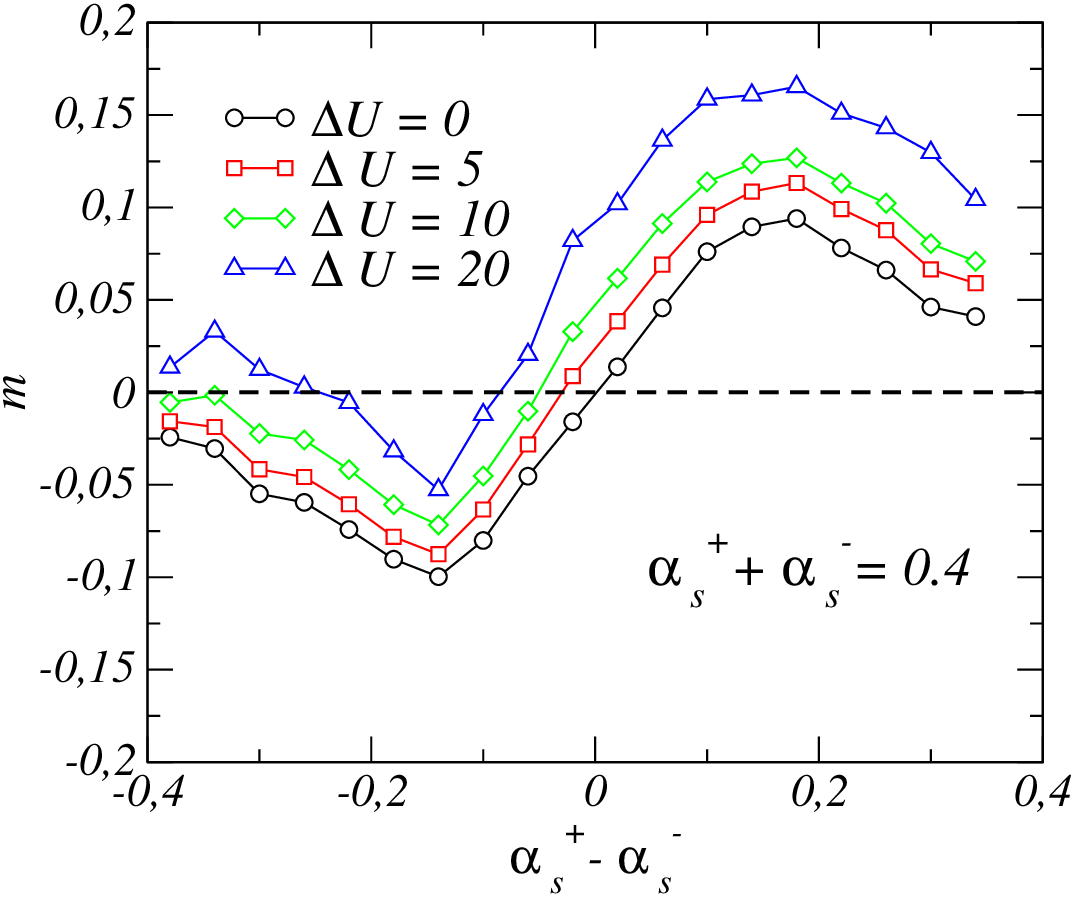}
\caption{Behavior of $m$ versus $\alpha_s^+-\alpha_s^-$ for
  $\alpha_s^++\alpha_s^-=0.4$, $\epsilon=0$, $n_p=1$ and biased
  initial conditions ($\Delta U=U_{i+}(0)-U_{i-}(0)$). Markers
  correspond to simulations with $N=200$ speculators, averaged over
  200 disorder samples per point.}
\label{initcon}
\end{figure}

Fig. \ref{initcon} shows the magnetization as
a function of $\alpha_+-\alpha_-$ along the cut
$\alpha_++\alpha_-=0.4$ in the phase diagram. This line is entirely contained
in the non-ergodic phase. While the market remains
globally predictable ($H>0$) the fact that one of the markets becomes
unpredictable (e.g. $H_+=0$) implies that the steady state depends on
initial conditions.
It is finally worth mentioning that the non-ergodic regimes with one
unpredictable market extend to large values of $\alpha_\gamma$.

\section{Conclusions}
\label{sect:concl}

We have studied a multi-asset version of the Minority Game in order to
address the problem of how adaptive heterogeneous agents would
diversify their investments when the different assets bear different
levels of information. While the phase structure of the models is
substantially a generalization of that of single-asset games, we have
found, in the grand-canonical model, a remarkable dependence of the
probability to invest in a certain asset on the agent's incentives
to trade ($\epsilon$).
Specifically, agents who have no incentives to trade other than the gains
derived from it, invest preferentially in
information-rich assets. On the contrary, when there are positive incentives to
trade ($\epsilon<0$) agents invest more likely in the information-poor asset.
This same behaviour is found in the canonical model, where agents must
choose one asset at each time step and cannot refrain from entering
the market.

The generalization of our results to a larger number of assets or to a wider
strategy pool for the agents is straightforward. The results discussed here are
indicative of the generic qualitative behavior we expect.

\begin{acknowledgements}
This work was supported by the European Community's Human Potential Programs
under contract COMPLEXMARKETS. 
\end{acknowledgements}


\begin{thebibliography}{}
\bibitem{ElFarol} W. B. Arthur, {\it Amer. Econ. Assoc. Papers and
Proc.} {\bf 84} 406 (1994).
\bibitem{CZ} D. Challet and Y.-C. Zhang, {\it Physica A} {\bf 246}
407 (1997).
\bibitem{book} D. Challet, M. Marsili and Y.-C. Zhang, {\it Minority
Games} (Oxford University Press, Oxford, 2005).
\bibitem{cool} A.C.C. Coolen, {\it The mathematical theory of Minority
Games} (Oxford University Press, Oxford, 2005).
\bibitem{traffic} A. De Martino, M.  Marsili and R. Mulet, {\it
Europhys. Lett.} {\bf 65} 283 (2004).
\bibitem{Johnson}
N. F. Johnson, P. M. Hui, D. F. Zheng and M. Hart, {\it J. of  Phys. A} {\bf 32}L427 (1999).
\bibitem{GCMG} 
D. Challet and M. Marsili, {\it Phys. Rev. E} {\bf 68}, 036132 (2003).
\bibitem{Mantegna} 
R. N. Mantegna, {\it Eur. Phys. Jour. B} {\bf 11}, 193 (1999).
\bibitem{Potters}
M. Potters, J. P. Bouchaud and L. Laloux,
cond-mat/0507111 (2005).
\bibitem{Kertesz}
 J. P. Onnela, A. Chakraborti, K. Kaski, J. Kertesz,
A. Kanto, {\it Phys. Rev. E} {\bf 68} (5) 056110 (2003).
\bibitem{Rodgers}
R. D'Hulst and G. J. Rodgers, adap-org/9904003 (1999).
\bibitem{Chau}
 F. K. Chow and H. F. Chau,
{\it Physica A} {\bf 319},601 (2003).
\bibitem{H} D. Challet, M. Marsili and R. Zecchina, {\it Phys. Rev.
    Lett.} {\bf 84} 1824 (2000)
\bibitem{MC01}
M. Marsili and D. Challet,
{\it Phys. Rev. E} {\bf 64},  056138  (2001).  
\bibitem{CoolHeim}J.A.F. Heimel and A.C.C. Coolen, {\it Phys. Rev. E} {\bf 63} 056121 (2001)  
\end{thebibliography}
\end{document}